\documentclass[useAMS,usenatbib,usegraphicx]{mn2e}
\usepackage{amsmath}
\usepackage[toc,page]{appendix}

	\usepackage{pslatex}
	\usepackage{amsmath}
\newcommand{\angstrom}{\mbox{\normalfont\AA}}
\title[ Dust Growth in Protoplanetary disks]{A Possible Mechanism for Overcoming the Electrostatic Barrier Against Dust Growth in Protoplanetary disks}
\author[V.V. Akimkin]{V.V. Akimkin$^{1,2}$
\\
$^{1}$Institute of Astronomy, Russian Academy of Sciences, Moscow, Russia \\
$^{2}$St. Petersburg State University, V.V. Sobolev Astronomical Institute, St. Petersburg, Russia
}
\begin{document}
%\centerline{\footnotesize   Received ; in final form,}\bigskip\noindent
\date{Received December 11, 2014. Accepted  January 22, 2015.  \\
akimkin@inasan.ru}
\pagerange{747--761} \pubyear{2015}
\volume{59}
\maketitle

%\author{\firstname{\bf V.~V.}~\surname{Akimkin}}
%
\begin{abstract}

The coagulation of dust particles under the conditions in protoplanetary disks is investigated.
The study focuses on the repulsive electrostatic barrier against growth of charged dust grains. Taking into
account the photoelectric effect leads to the appearance of a layer at intermediate heights where the dust has
a close to zero charge, enabling the dust grains to grow efficiently. An increase in the coagulation rate
comes about not only due to the lowering of the Coulomb barrier, but also because of the electrostatic
attraction between grains of opposite charge due to the non-zero dispersion of the near-zero charge.
Depending on the efficiency of mixing in the disk, the acceleration of the evolution of the dust in this layer
could be important, both in the quasi-stationary stage of the disk evolution and during its dispersal.

DOI: 10.1134/S1063772915070021
\end{abstract}  

%\maketitle 

\def\gtrsim{\mathrel{\hbox{\rlap{\hbox{\lower5pt\hbox{$\sim$}}}\hbox{$>$}}}}

\section{Introduction}

The first attempts to describe the growth of particles
due to their coagulation were made by
Smoluchowski~\citep{1916ZPhy...17..557S} in 1916. He presented an equation
(the so-called Smoluchowski equation) describing
the agglomeration of colloidal particles undergoing
Brownian motion. The methods of the coagulation
theory were adapted to the study of processes in the
late, gravity-assisted accumulation of macroscopic pre-planetary bodies by Safronov~\citep{1969edo..book.....S}. Subsequently, these methods have also been widely
applied to theoretical modeling of the early stages of
the evolution of protoplanetary disks~\citep{1997A&A...325..569S,2005ApJ...625..414T,2006ApJ...640.1099N,2010A&A...513A..79B,2013ApJ...766....8A}. 

In spite of successes in theoretical modeling of the
evolution of dust in protoplanetary disks~\citep{2014arXiv1402.1354T}, many
important problems remain to be solved.
These include
so-called growth barriers, which hinder
the formation of macroscopic dust particles~($\gtrsim1$\,cm in size). The first of these barriers (arising earliest in
the evolution of the disk) is the electrostatic repulsion
between dust grains with the same sign of charge~\citep{2009ApJ...698.1122O, 2012ApJ...744....8M}. Due to the negative charge of the grains in deep layers of the disk, an appreciable suppression of the dust coagulation can arise, even causing
this process to fully cease. As was noted in~\cite{2011ApJ...731...96O}, overcoming the Coulomb barrier is possible under
the conditions of strong turbulence ($\alpha\gtrsim10^{-2}$), but
this can have an adverse effect on the growth of
larger grains due to fragmentation during collisions. Overall, possible ways to overcome the electrostatic
barrier have been relatively little studied, compared to
barriers to dust growth arising at later stages in the
evolution of protoplanetary disks.

Although the charge of dust is an important (if not
the determining) factor in its dynamics and evolution,
the coagulation of even neutral dust grains can
be suppressed by various mechanisms. For example,
grains become more compact due to collisions, which increases the probability that they will bounce, rather
than coagulate. This leads to a reduction in the
growth of the grains and the appearance of a so-called
 bouncing barrier~\citep{2010A&A...513A..57Z}. As the dust grains grow,
 their relative velocity increases due to their radial drift~\citep{1977MNRAS.180...57W} and turbulence induced motion~\citep{1980A&A....85..316V}. If the collision velocity exceeds
 some critical value ($\sim 10$\,m/s), the collisions likely lead to fragmentation, rather than agglomeration, of
 the grains --- the so-called fragmentation
barrier~\citep[]{2008ARAA..46...21B, 2009ApJ...702.1490W, 2009MNRAS.393.1584T}.
 Radial drift can remove large grains from the disk
so quickly that they simply are not able to grow during
their motion toward the central star~\citep{2008A&A...480..859B}. 
This imposes
strong lower limits on the rate of coagulation
of the particles in protoplanetary disks at the final
stages of evolution (the radial-drift barrier).

Only collisional charging is usually considered in
modeling of the coagulation of charged dust
grains~\citep{2009ApJ...698.1122O,2011ApJ...731...96O, 2012ApJ...744....8M}. In this case, under the conditions near
the midplane of a protoplanetary disk, nearly
all the grains are negatively charged. However, in
the atmosphere of the disk, where the photoelectric
effect becomes important, the dust grains are instead strongly positively charged~\citep{2011ApJ...740...77P}. Accordingly, there
must exist a region at intermediate heights where the
grains have zero charge, or even opposite charges
(due to the dispersion of the charge in the ensemble
of grains), which can potentially appreciably increase
the coagulation rate. In our study, we investigated
the possibility and conditions for the growth
of charged dust in a layer of grains with near-zero
charge for the typical conditions of a protoplanetary
disk.

Note that numerous interesting effects in dusty
plasmas have been discovered in experiments under
microgravity conditions on-board the Mir space station~\cite{2004PhyU...47..447F}, and also subsequently on-board the International
Space Station. In particular, an anomalous
increase in the rate of coagulation of charged
grains (by four to five orders of magnitude) was found,
compared to the case of neutral grains~\citep{2005NJPh....7..227K}.

The structure of the paper is as follows. Section 2
describes the method used, and a discrete equation for
the coagulation of charged grains with a non-zero
charge dispersion is presented. Section 3 presents
the results of dust grain coagulation modeling in a protoplanetary disk for neutral dust
and two cases of charged dust. Section 4 discusses
dynamical aspects of the evolution of grains and the
extent to which the adopted assumptions influence
the modeling results. The Conclusion summarizes
the main conclusions of our study.

\section{COAGULATION THEORY APPLIED TO THE GROWTH OF CHARGED GRAINS}

\subsection{General Form of the Coagulation Equation}

The Smoluchowski equation can be written in two
forms~--- integral and discrete. The former considers
the continuous distribution of the particle masses, $f(m)$ [cm$^{-3}$ g$^{-1}$], where $f(m)dm$ is the number density
of particles with masses within $dm$ of $m$. The
distribution $f(m)$ is normalized to the total number
density of dust particles $n_d=\int f(m)dm$. The coagulation rate for particles with masses $m_1$ and $m_2$ is specified by the coagulation kernel $K(m_1,m_2)$, which is proportional to the collisional cross section, the velocity of the colliding particles, and the sticking probability. In the integral form, the evolution of the distribution function $f(m)$ is given by the equation
\begin{align}
 \frac{\partial f(m,t)}{\partial t}&=\frac{1}{2}\int\limits_0^m f(m-m',t) f(m',t) K(m-m',m') dm'-\nonumber \\
 &-f(m,t)\int\limits_0^{\infty}f(m',t)K(m,m')dm'.
\end{align}
The first term here describes the coagulation of particles
with masses $m-m'$ and $m'$, leading to the
formation of a particle with mass $m$. The second term
is negative and corresponds to the disappearance of
particles with mass $m$ due to their coagulation with
all other particles. An analytical solution to this equation
can be found only for the simplest coagulation
kernels~--- constant~\citep{1916ZPhy...17..557S} and linear~\citep{1969edo..book.....S}. These kernels
are not suitable for the conditions in a protoplanetary
disk, but can be used to test algorithms for the solution
of the Smoluchowski equation. A numerical
approach can be used with more realistic kernels.
A solution of this equation for the conditions in a
protoplanetary disk is presented, for example, in~\cite{2005A&A...434..971D}.

The discrete form of the coagulation equation
describes the variations in the number densities $\{N_k\}_{k=1}^{\infty}$ of particles with masses $\{m_k\}_{k=1}^{\infty}$ proportional
to the minimum mass of a monomer $m_0$: $m_k=km_0$. In this case, the coagulation equation
takes the form~(see, e.g., the review \cite[Chapter III,
Section 6]{1943RvMP...15....1C}):

\begin{align}
 \dot{N_k}=\frac{1}{2}\sum_{i+j=k}N_iN_jK_{ij}-N_k\sum_{j=1}^{\infty}N_jK_{kj}.
\end{align}
Note that only terms in which $m_i+m_j=m_k$ are
included in the first (double) sum, which is equivalent
to the condition $i+j=k$ for the adopted mass grid. Finding an exact solution to this equation requires
tracing a large range of particle masses, which makes
the grid $m_k=km_0$ unsuitable in practice. Therefore, a
logarithmic grid of masses of the coagulating particles
is usually used. The problem of conservation of
mass arises, since two arbitrary masses from such a
grid $m_i$ and $m_j$ are very unlikely to correspond to a $k$ such that $m_k=m_i+m_j$. A conservative algorithm
is presented in~\cite{KovetzOlund1969}, in which the mass $m_i+m_j$ distributed between the masses for the nearest grid
points to the left and right $m_l$ and $m_r$. The discrete
coagulation equation can then be written in the form
\begin{align}
 \dot{N_k}=\frac{1}{2}\sum_{i=1}^{k}\sum_{j=1}^{k}N_iN_jC_{ijk}K_{ij}-N_k\sum_{j=1}^{\infty}N_jK_{kj},\label{discretecoag}
\end{align}

\begin{equation}
C_{ijk}=\begin{cases}
  \epsilon,&\text{if } m_k=\max\limits_{m_n<m_i+m_j}\{m_n\};\\
  1-\epsilon,&\text{if } m_k=\min\limits_{m_n>m_i+m_j}\{m_n\};\\
  0,&\text{otherwise,}
\end{cases}
\end{equation}

\begin{equation}
 \epsilon=\frac{m_r-(m_i+m_j)}{m_r-m_l}.
\end{equation}

Even in the conservative form~\eqref{discretecoag}, the coagulation
equation is complicated to solve. This is related to
the limited length of the mantissa of the computer representation of a real number. The coagulation
of particles with masses differing by 15 orders of
magnitude will be comparable to the rounding errors,
even using double-precision variables in the computations.
Under the conditions of protoplanetary disks,
the range of dust grain masses can reach 20 orders
of magnitude or more. Therefore, effective (but unwieldy)
algorithms have been developed, that make it
possible to include a wide range of masses without the
need for a transition to high-precision arithmetical
computations~\citep{2008A&A...480..859B}. For the goals of our study, it
is important to preserve the simple form~\eqref{discretecoag}; therefore,
we chose an extensive approach based on using
libraries designed for operations with real numbers
with arbitrary mantissa lengths. The rising computational
expenses for this case were compensated by the application of parallel-programming methods.

\subsection{Coagulation Equation Kernel with Grain Charge Dispersion}

The efficiency of the coagulation process is described
by the kernel $K_{ij}=K(m_i,m_j)$. The case
of neutral grains has been studied fairly well recently
(see, e.g., \cite{2008A&A...480..859B,2010A&A...513A..79B,2010A&A...513A..57Z,2008ARAA..46...21B}):
\begin{equation}
 K_{ij}=p_{ij}\pi(a_i+a_j)^2u_{ij},
\end{equation}
where $p_{ij}$ is the sticking probability for grains with
radii $a_i$ and $a_j$ moving with relative velocity $u_{ij}$. We
assumed that the relative velocity of the grains arises
only due to Brownian motion, which corresponds to
the initial stages of dust evolution (and grain sizes
up to $\sim10^{-2}$\,cm at distances of 1 AU; see, e.g., \cite{2014arXiv1402.1354T}). Other mechanisms determining the relative velocities
of the grains, such as turbulence and radial and vertical
drift~\citep{2008A&A...480..859B} become important at later stages of dust
evolution than those considered here. Including these
in models for charged dust requires a much more
complex approach and is the topic of a separate study.

The charges of grains in the central regions of a
disk were calculated in detail in~\cite{2009ApJ...698.1122O,2011ApJ...731...95O,2011ApJ...731...96O}, where the
electrostatic interaction energy for large grains (with
sizes of more than $10^{-3}$\,cm) was found to exceed
their kinetic energy associated with collisions~--- the
electrostatic barrier against dust growth referred to
above.

To analyze the influence of the grain charge on
the rate of their growth in more detail, we considered
a general form of the coagulation kernel taking into
account Coulomb focusing~\citep{1941ApJ....93..369S}:

\begin{equation}
 K_{ij}(Q_i,Q_j)=p_{ij}\pi(a_i+a_j)^2\left[1-\frac{2Q_iQ_je_p^2}{m_{ij}(a_i+a_j)u_{ij}^2}\right]u_{ij}, \label{CoulombCross}
\end{equation} 
where $Q_i$ and $Q_j$ are the charges of the grains, $m_{ij}=m_im_j/(m_i+m_j)$ is the reduced mass, and $e_p$ is the
elementary charge. It is obvious that the additional
condition  $K_{ij}(Q_i,Q_j)\ge0$ is imposed on $K_{ij}$. This
form of the coagulation kernel~\eqref{CoulombCross} is valid when the
mass and size of the grains uniquely determine their
charge.

However, the charge of a specific grain fluctuates
with time; in other words, there is a non-zero dispersion
for the charge in an ensemble of grains of a given
size. Let the mean charge of grain $i$ be $ \overline{Q}_i$ and the
standard deviation of the charge about the mean be
$\sigma_i$. As a rule~\citep{2005pcim.book.....T}, the distribution of grains of a given size $a_i$ over the charge $q$ will be close to normal
\begin{equation}
f(q,\overline{Q}_i,\sigma_i)= \frac{1}{\sigma_i\sqrt{2\pi}}e^{-\frac{(q-\overline{Q}_i)^2}{2\sigma_i^2}}.
\end{equation}
 It is not difficult to show that the coagulation kernel
transforms to the form

\begin{equation}
 K^*_{ij}=\int f(q,\overline{Q}_i,\sigma_i) \int f(p,\overline{Q}_j,\sigma_j) K_{ij}(q,p) \,dq \,dp. \label{charged_kernel}
\end{equation}
The simple replacement of $K_{ij}$ with $K^*_{ij}$ brings Eq.\eqref{discretecoag} for neutral grains to the discrete coagulation equation
for charged grains with a non-zero charge dispersion.
The sticking probability was taken to be unity. The
process of fragmentation during collisions at high velocities
is also important for the evolution of dust. The
fragmentation barrier against dust growth is important
for large grains; i.e., it becomes significant after
the electrostatic barrier has been overcome. Since our
aim was to search for mechanisms capable of overcoming
the electrostatic barrier, we neglected grain
fragmentation.

Equation~\eqref{discretecoag} written for each $k=1,\dots,N_{\rm mass}$ represents a system of non-linear ordinary differential
equations. We solved this system using an explicit
5th-order Runge--Kutta method with an adaptive
step control~\citep{Press:1993:NRF:563041}. Although implicit integration
methods can yield a solution over a smaller number of
steps, in practice, such methods proved slower due to
the additional time required to compute the Jacobian.

\subsection{Computation of the Dust Charge}

The photoelectric effect and the accretion of electrons
and ions onto a grain are the main factors determining
the charge of a dust grain under most astrophysical
conditions \citep{2001ApJS..134..263W}. Thus far, only the acquisition
of charge through collisions has been considered
in relation to the dust growth in protoplanetary disks.
In this case, the high velocities of electrons compared
to ions lead to negatively charged grains. However, as a rule (see, e.g.,\citep{2011ApJ...740...77P}, a positive charge may be
acquired in the atmosphere of the disk due to the
efficient photoelectric effect. Therefore, there should
exist a region in the disk at intermediate heights
where the grain charges are close to zero. The main
goal of our study was a detailed consideration of this
region from the point of view of the coagulation theory
of charged particles. The main equations used to
compute the grain charges, based on the treatment
in~\cite{2005pcim.book.....T,1987ApJ...320..803D,2001ApJS..134..263W}, are presented below.

If we consider collisions of dust grains with electrons
and singly charged ions, the probability $f(Z_{\rm gr})$ of encountering a grain with charge $Z_{\rm gr}e_p$ can be
found from the detailed balance equation

\begin{equation}
 f(Z_{\rm gr})\left[J_{\rm pe}(Z_{\rm gr})+J_{\rm ion}(Z_{\rm gr})\right]  =  f(Z_{\rm gr}+1)J_{\rm e}(Z_{\rm gr}+1),
\end{equation}
where $J_{\rm pe}$ [s$^{-1}$] is the photoelectron emission rate and $J_{\rm e}$ and $J_{\rm ion}$ [s$^{-1}$]are the accretion rates of electrons
and ions. To find the total charge distribution function  $f(Z_{\rm gr})$, the equation above is written for each value $Z_{\rm gr}$, which reduces to the system
 \begin{equation} 
  f(Z_{\rm gr})=f_0\prod\limits_{z=Z_0+1}^{Z_{\rm gr}}\frac{J_{\rm pe}(z-1)+J_{\rm ion}(z-1)}{J_{\rm e}(z)} \label{fZ1}
 \end{equation}
for $Z_{\rm gr}>Z_0$ and
\begin{equation}
 f(Z_{\rm gr})=f_0\prod\limits_{z=Z_{\rm gr}}^{Z_0-1}\frac{J_{\rm e}(z+1)}{J_{\rm pe}(z)+J_{\rm ion}(z)}  \label{fZ2}
\end{equation}
for $Z_{\rm gr}<Z_0$. Here, $f_0\equiv f(Z_0)$. We especially note
that, from a computational point of view, it is important
to choose a value of $Z_0$ that is close to the maximum
of $f(Z_{\rm gr})$, which can be tentatively estimated
from the approximate equality
\begin{equation}
 J_{\rm pe}(Z_0)+J_{\rm ion}(Z_0)\approx J_{\rm e}(Z_0). \label{Zapp}
\end{equation}
Equations \eqref{fZ1}, \eqref{fZ2} are closed with the obvious relation
\begin{equation}
 \sum\limits_{Z=-\infty}^{\infty} f(Z) =1.
\end{equation}

The accretion rate of particles with number density
 $n$ and mass $m$ can be written~\citep{1987ApJ...320..803D}:
\begin{equation}
 J_{\rm acc}=n\sqrt{\frac{8kT_{\rm gas}}{\pi m}}\pi a_{\rm gr}^2 \tilde{J}(a_{\rm gr},T_{\rm gas},Z_{\rm gr}).
\end{equation}
The function $\tilde{J}$ takes into account Coulomb focusing.
Following~\cite{2001ApJS..134..263W}, we took the sticking probabilities for
electrons and ions to be 0.5 and 1, respectively.

The photoemission rate depends on the
solid-angle-averaged radiation intensity $J_{\nu}$ [erg/(cm$^2$ s Hz ster)], the absorption cross
section $C_{\rm abs}$ and the photo-ionization yield $Y$~\citep{2001ApJS..134..263W}:
\begin{equation}
 J_{\rm pe}=\int \frac{4\pi J_{\nu}}{h\nu}C_{\rm abs}(\nu,a_{\rm gr}) Y(\nu,Z_{\rm gr},a_{\rm gr})\,d\nu
\end{equation}
An important assumption we made is that the grains
are not the dominant charge carriers in the plasma.
The depth of penetration of the photons and the mean
free path of the electrons in a grain were taken to be 100\,\AA\, and 10\,\AA, respectively, and the work function of
the photoelectrons to be 8~eV, which corresponds to
silicate grains.

\section{MODELING THE GROWTH
OF CHARGED GRAINS}

To show how the charge of a grain influences
its growth, the coagulation equation~\eqref{discretecoag} was solved
locally for each point in the disk for three cases. In
the first, the grains were assumed to be neutral; i.e.,
their coagulation is not hindered (or facilitated) by
electrostatic interactions. We call this case the comparison
model (Model 1). In the second (Model 2), it
is assumed that the charge of a grain is determined
only by collisions with electrons and ions. This approximation
is valid for dense regions in the disk near
the midplane, where stellar radiation cannot reach.
In this case, the grains become negatively charged,
and the Coulomb interaction begins to hinder coagulation.
Both grain-charging mechanisms are considered
in Model 3~--- collisional and radiative. This
general case is of the most interest.

We emphasize that the coagulation equation was
solved locally and independently for each point in the
disk. In other words, it was assumed that there was
no advective term in the coagulation equation. This
assumption is very probably not satisfied for large
grains in protoplanetary disks, since they are subject
to settling toward the midplane and radial drift. However,
such a local approach is useful when searching
for possible ways to enhance the growth of grains in
some part of the disk. In this case, the distribution of
the mean grain size over the disk visually reflects the
distribution of coagulation rates.

\subsection{Model for the Physical Structure of the Disk}

Since we did not aim to compute the detailed
physical structure of the disk (density, temperature,
radiation field), we chose the characteristic parameters
of disks around single T Tauri stars~\citep{2011ARA&A..49...67W}.  The disk
was assumed to be stationary up to an age of one to
three million years~\citep{2009AIPC.1158....3M}, with a surface density $\Sigma$, that
has a power-law dependence on the distance to the
star $R$: $\Sigma=\Sigma_0 (R/R_0)^{p_{\rm s}}$. We assumed hydrostatic
equilibrium in the vertical direction. The radial profile of the temperature was determined assuming a geometrically
thin disk and blackbody re-radiation of the
UV flux from the star \cite[formula (2.28)]{2010apf..book.....A}. The disk
was taken to be isothermal in the vertical direction,
and the gas temperature was taken to be equal to the
dust temperature. The stellar radiation was taken to
be blackbody radiation with the effective temperature $T_{\star}$ and a UV excess with effective temperature $T_{\star}^{\rm UV}$. We chose the normalization of the UV excess such
that $B_{4000\angstrom}(T_{\star})=B_{4000\angstrom}(T_{\star}^{\rm UV})$. The degree of ionization
was taken to be the same throughout the disk and equal to $x_{\rm e}=10^{-4}$, which corresponds to the
ionization of carbon (we discuss the influence of this
assumption for the model of the dust evolution below).
The coagulation equation was solved locally for
each point in the disk on a $N_{\rm r}\times N_{\rm z}$ grid.

The influence of the assumption of stationarity and
the choice of disk parameters on the results obtained
is discussed in Section~\ref{neUV}. The results of our computations
for Models 1--3 with the disk parameters listed
in the Table 1 are presented below.%~\ref{Tab_params}.

%%%%%%%% Table 1 moved to the end of manuscript %%%%%%%%%%
%Table 1
\begin{table}
\begin{center}
\caption{Main parameters of the model}
\begin{tabular}{c|c}
\hline
 Parameter & Value      \\
 \hline
  $R_0$ &1 AU \\ 
 $\Sigma_0$& 300 g/cm$^2$\\
 $p_{\rm s}$ &-1  \\
 $M_{\rm disk}$ &0.02  $M_{\odot}$ \\
 $T_{\star}$ & 4000 K \\
 $T_{\star}^{\rm UV}$ & 20000 K \\
 $x_e$ & $10^{-4}$  \\
 $N_{\rm mass}$ & 64--128 \\
 $N_{\rm r}$ & 15--30  \\
 $N_{\rm z}$ & 15--30 \\
  \hline
\end{tabular} \label{Tab_params}
\end{center}
\end{table}
%%%%%%%% Table 1 moved to the end of manuscript %%%%%%%%%%

%%%%% Section 3.1
\subsection{Neutral Grains}

\begin{figure}
%\setcaptionmargin{5mm}
%\captionsetup{singlelinecheck=false,justification=raggedright}
\includegraphics[width=0.49\textwidth]{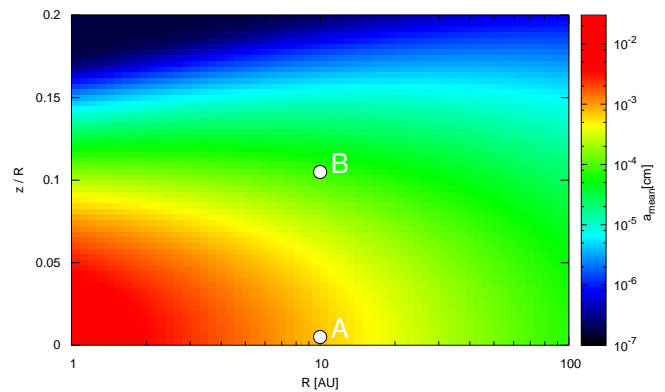} 
\caption{Distribution of the mean grain size over the disk for Model 1 at a time of three million years.}
\label{Fig_A_null}
\end{figure}

The case of neutral grains has been considered
in many studies of dust evolution in protoplanetary
disks. Excluding Coulomb forces from consideration
leads to faster grain growth for denser regions. Since
the coagulation rate is proportional to the square of
the particle number density, the mean size of the
grains $\overline{a}$ depends strongly on their location in the
disk~(Fig.~\ref{Fig_A_null}). 
%% Fig 1 (A_null_gray) was moved to the end of manuscript
%Fig.1
%% Fig 1 (A_null_gray) was moved to the end of manuscript
Three million years after the start of
the growth, the range of the mean grain sizes in the
disk reaches six orders of magnitude (18 order of
magnitude in mass). Note that fragmentation can
appreciably slow the dust growth rate, thereby narrowing
this range. However, as was shown in~\cite{2012A&A...544L..16W}, taking into account the velocity distribution of the
grains partially solves the problem of fragmentation.

\begin{figure}
%%\setcaptionmargin{5mm}
%\captionsetup{singlelinecheck=false,justification=raggedright}
\includegraphics[angle=270,width=0.49\textwidth]{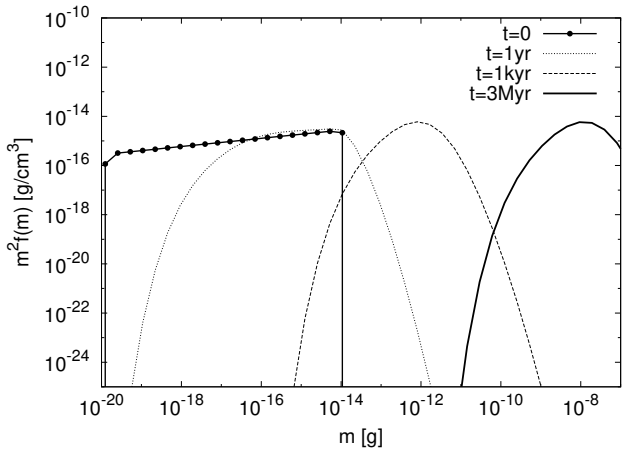} \\
\includegraphics[angle=270,width=0.49\textwidth]{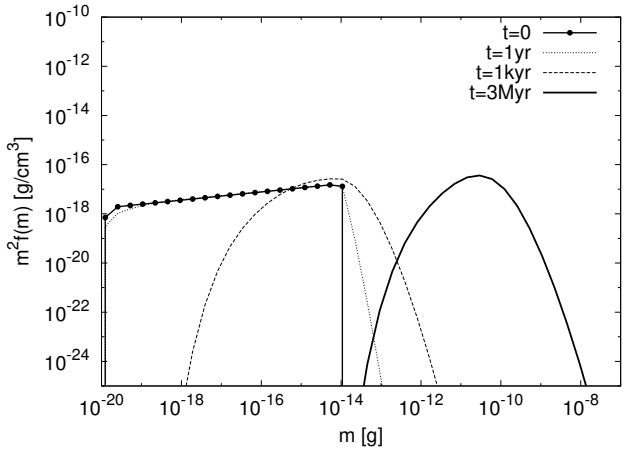} 
\caption{Mass distribution function for the grains for Model 1 (for various times) at point A (upper panel) and point B (lower panel) marked in Fig.~\ref{Fig_A_null}.}
\label{Fig_ff_null}
\end{figure}

\begin{figure}
%\setcaptionmargin{5mm}
%\captionsetup{singlelinecheck=false,justification=raggedright}
\includegraphics[angle=270,width=0.49\textwidth]{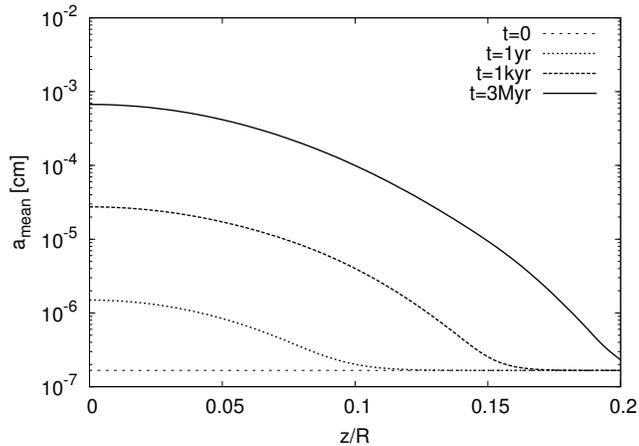} 
\caption{Vertical profile of the mean grain size at a distance of 10 AU from the star for Model 1 (for various times).}
\label{Fig_vertical_null}
\end{figure}

\begin{figure}
%\setcaptionmargin{5mm}
%\captionsetup{singlelinecheck=false,justification=raggedright}
\includegraphics[angle=270,width=0.49\textwidth]{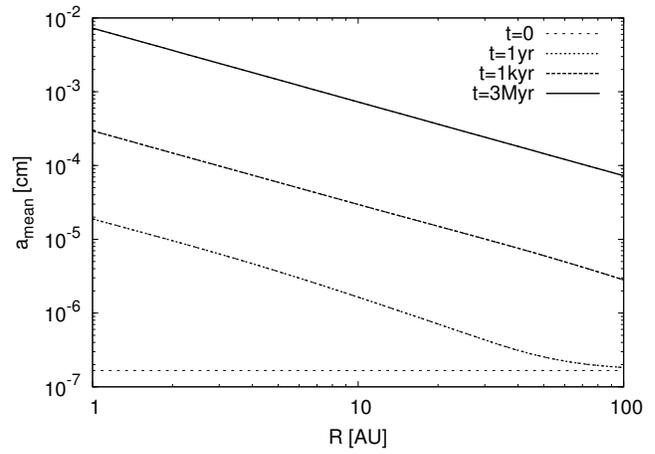} 
\caption{Radial profile of the mean grain size in the midplane of the disk for Model 1 (for various times).}
\label{Fig_radial_null}
\end{figure}

The detailed mass distributions of the grains $f(m)$ for points A and B indicated in Fig.~\ref{Fig_A_null} at a radius of
10 AU and various times are also presented (Fig.~\ref{Fig_ff_null}). The densities at points A and B differ by about three orders of magnitude. Figure~\ref{Fig_ff_null} presents the quantity $m^2f(m)$, for Model 1, which reflects the contributions
of various grains to the total mass. 
%% Fig 2 (ff1_null_gray and ff16_null_gray) was moved to the end of manuscript
%Fig.2
%% Fig 2 (ff1_null_gray and ff16_null_gray) was moved to the end of manuscript
 The zero of time
corresponds to a so-called MRN distribution~\citep{1977ApJ...217..425M}. The growth of the dust at points A and B occurs
similarly in terms of the profiles of the distribution
function $f(m)$, but on different time scales. Figures \ref{Fig_vertical_null} and \ref{Fig_radial_null} 
%% Fig 3 (vertical_null_gray) was moved to the end of manuscript
%Fig.3
%% Fig 3 (vertical_null_gray) was moved to the end of manuscript
%% Fig 4 (radial_null_gray) was moved to the end of manuscript
%Fig.4
%% Fig 4 (radial_null_gray) was moved to the end of manuscript
show the vertical and radial profiles of the mean
grain size for times $t=1,10^3$ and $3\cdot10^6$~yr. Note that the growth of neutral grains is fairly uniform
in the sense that $\lg \overline{a}\sim \lg t$, which is not true for
charged grains.

%%%%%%%%%%%%%%%%%%%%%%%%%%%%%%%%%%%%%%%%%%%%%%%%%%%%%%%%%%%%%%%%%%%%%%%%%%%%%%%%%%%%%%%%%%%%%%%%%%%%%%%%%%%%%%%%%%%%%%%%%%% Section 3.2
\subsection{Collisionally Charged Grains}

Under equilibrium conditions, the electrons have
higher velocities than the ions, so that the grains
collide more often with electrons, leading to a negative
charge for the grains. As the grains grow, their
negative charge also increases. Their relative velocity
may also increase. Therefore, the Coulomb factor in Eq.~\eqref{CoulombCross} can either increase or decrease during the dust
evolution. It was shown in~\cite{2009ApJ...698.1122O} that the electrostatic
interaction energy begins to exceed the mean kinetic
energy of grains with sizes of $\sim 10^{-3}$\,cm or more, so that the growth of the grains should be appreciably
suppressed. This threshold should depend on the
ambient conditions, which can vary strongly over the
disk. Therefore, we carried out global modeling of the
coagulation of (negatively) charged grains in order to
study the conditions for the appearance of the electrostatic
barrier.

\begin{figure}
%\setcaptionmargin{5mm}
%\captionsetup{singlelinecheck=false,justification=raggedright}
%\includegraphics[angle=270,width=0.49\textwidth]{./Figures/A_coll_gray}
\includegraphics[width=0.49\textwidth]{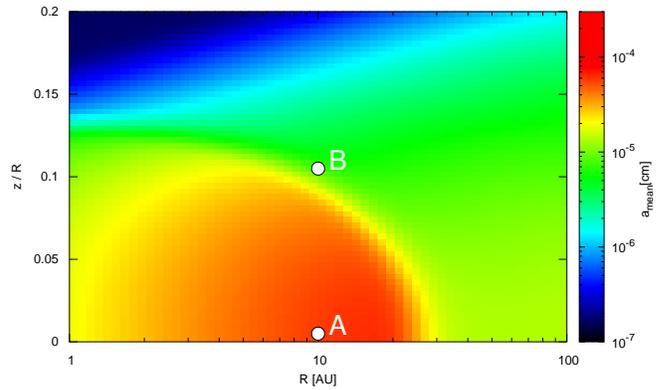} 
\caption{Same as Fig.~\ref{Fig_A_null} for Model 2. The upper boundary of the color scale is two orders of magnitude lower than the scale in Fig.~\ref{Fig_A_null}.} 
\label{Fig_A_coll}
\end{figure}

Figure~\ref{Fig_A_coll} presents the distribution of the mean
grain size over the disk, taking into account collisional
charging. The growth of the dust is appreciably
suppressed by its charge, and the upper boundary for
the grain size shown by the color scale is two orders of
magnitude lower than in Fig.~\ref{Fig_A_null}.
%% Fig 5 (A_coll_gray) was moved to the end of manuscript
%Fig.5
%% Fig 5 (A_coll_gray) was moved to the end of manuscript
 The main difference
from the case of neutral dust is that the growth rate
is not proportional to the number density. This is expressed
by the fact that the grains grow most rapidly
at intermediate radii $\approx2-20$~AU. The grains are more
strongly charged in inner, dense regions than in outer
regions, making the electrostatic suppression of their
growth more important there. With distance from
the star, the grain charge begins to decrease in magnitude,
speeding up the coagulation. The drop in
density in the disk becomes important in peripheral
regions, leading to slowing of the coagulation.
 
\begin{figure}
%\setcaptionmargin{5mm}
%\captionsetup{singlelinecheck=false,justification=raggedright}
\includegraphics[angle=270,width=0.49\textwidth]{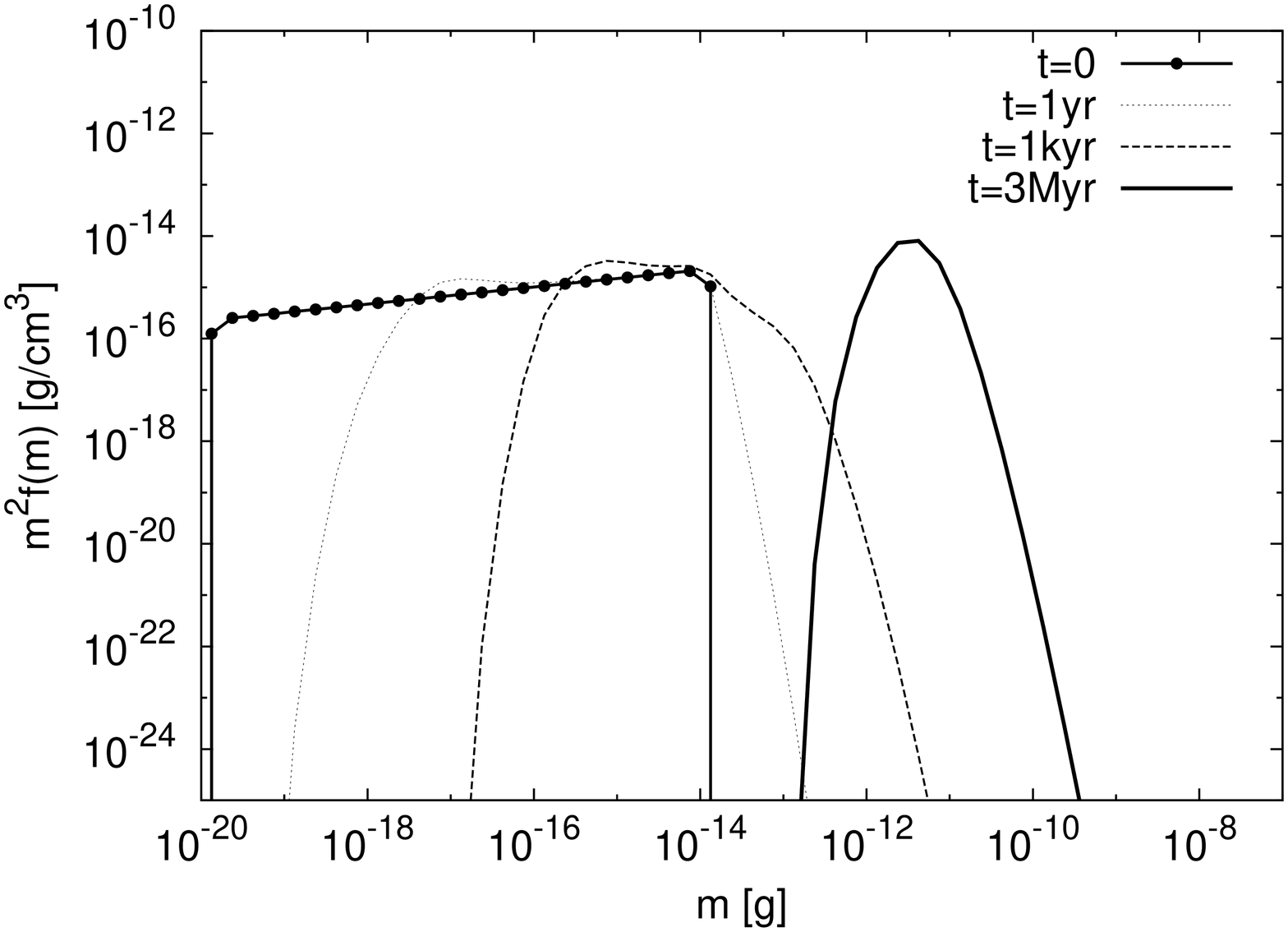} \\
\includegraphics[angle=270,width=0.49\textwidth]{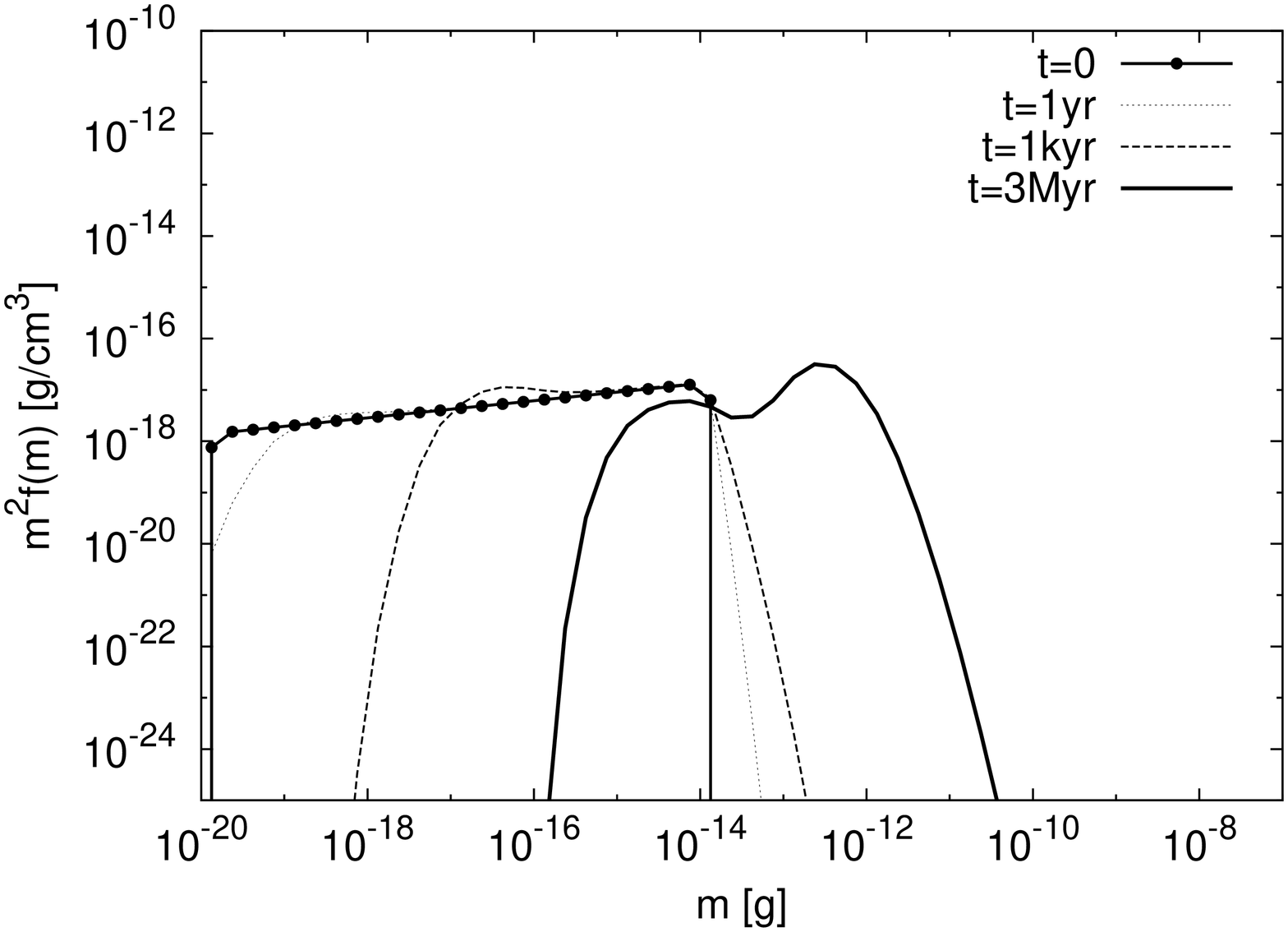} 
\caption{Same as Fig.~\ref{Fig_ff_null} for Model 2.}
\label{Fig_ff_coll}
\end{figure}

The mass distribution function of the charged
grains also differs qualitatively from the neutral case (Fig.~\ref{Fig_ff_coll}). This distribution is bimodal; this is clearly
visible for point B at a time of three million years,
but is also true to a lesser extent for other times and
locations in the disk. The origin of this bimodality is
related to the non-geometrical relationship between the interaction cross section and the particle size. If
the mass growth rate was proportional to the geometrical
cross section, particles with initially different
sizes would tend to acquire more equal sizes as they
grew. Additional factors, whether gravitational or
electrostatic forces, change the effective collisional
cross section. This means that large particles can
grow more rapidly than smaller ones.
%% Fig 6 (ff1_coll_gray and ff16_coll_gray) was moved to the end of manuscript
%Fig.6
%% Fig 6 (ff1_coll_gray and ff16_coll_gray) was moved to the end of manuscript

\begin{figure}
%\setcaptionmargin{5mm}
%\captionsetup{singlelinecheck=false,justification=raggedright}
\includegraphics[angle=270,width=0.49\textwidth]{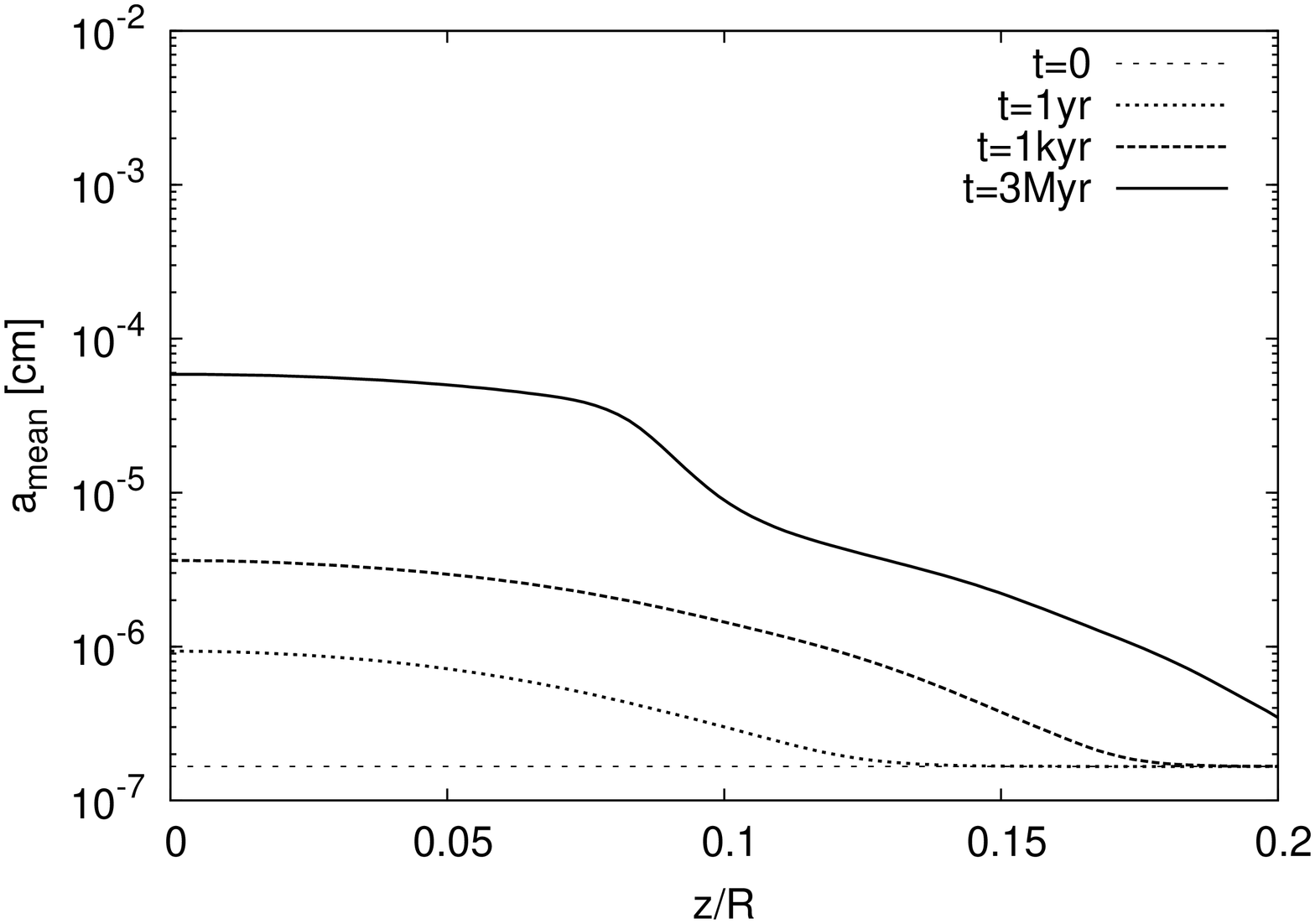} 
\caption{Same as Fig.~\ref{Fig_vertical_null} for Model 2.}
\label{Fig_vertical_coll}
\end{figure}

\begin{figure}
%\setcaptionmargin{5mm}
%\captionsetup{singlelinecheck=false,justification=raggedright}
\includegraphics[angle=270,width=0.49\textwidth]{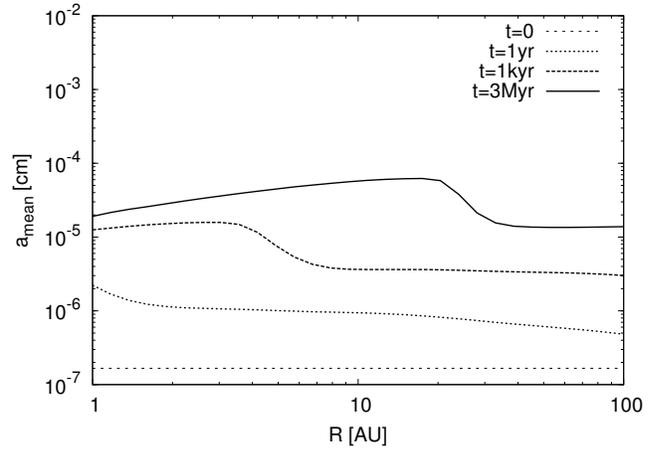} 
\caption{Same as Fig.~\ref{Fig_radial_null} for Model 2.}
\label{Fig_radial_coll}
\end{figure}

 A comparison of the vertical and radial profiles of
the mean grain size for Model 2 (Figs.~\ref{Fig_vertical_coll} and \ref{Fig_radial_coll}) and Model~1 (Figs.~\ref{Fig_vertical_null} and \ref{Fig_radial_null}) confirms the need to take
into account the dust charge when considering its
evolution. 
%% Fig 7 (vertical_coll_gray) was moved to the end of manuscript
%Fig.7
%% Fig 7 (vertical_coll_gray) was moved to the end of manuscript
%% Fig 8 (radial_coll_gray) was moved to the end of manuscript
%Fig.8
%% Fig 8 (radial_coll_gray) was moved to the end of manuscript
 Note that the maximum of the profiles in Fig.~\ref{Fig_radial_coll} move outward during the evolution of the dust.

%%%%%%%%%%%%%%%%%%%%%%%%%%%%%%%%%%%%%%%%%%%%%%%%%%%%%%%%%%%%%%%%%%%%%%%%%%%%%%%%%%%%%%%%%%%%%%%%%%%%%%%%%%%%%%%%%%%%%%%%%%% Section 3.3
\subsection{General Case of Charged Dust}

The ejection of electrons from grains by UV photons
leads to appreciably positively charged dust in
the disk atmosphere. As the UV flux weakens toward
the midplane, the efficiency of the photoelectric effect
decreases, and the grains become less charged. At
some height above the midplane, the mean grain
charge becomes zero.
%% Fig 9 (Q_full_gray and sigma_full_gray) was moved to the end of manuscript
%Fig.9
\begin{figure}
%\setcaptionmargin{5mm}
%\captionsetup{singlelinecheck=false,justification=raggedright}
\includegraphics[angle=270,width=0.49\textwidth]{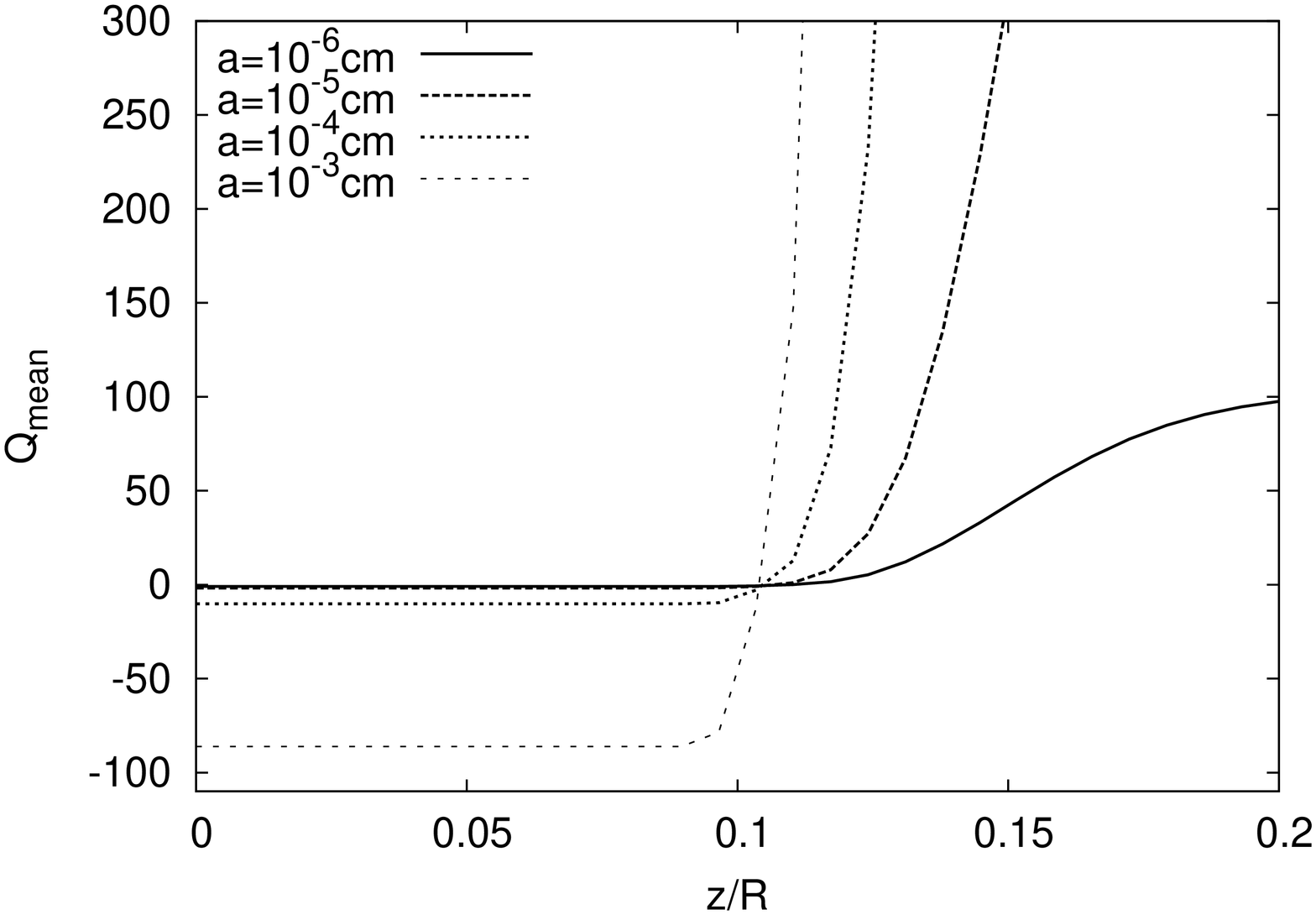} \\
\includegraphics[angle=270,width=0.49\textwidth]{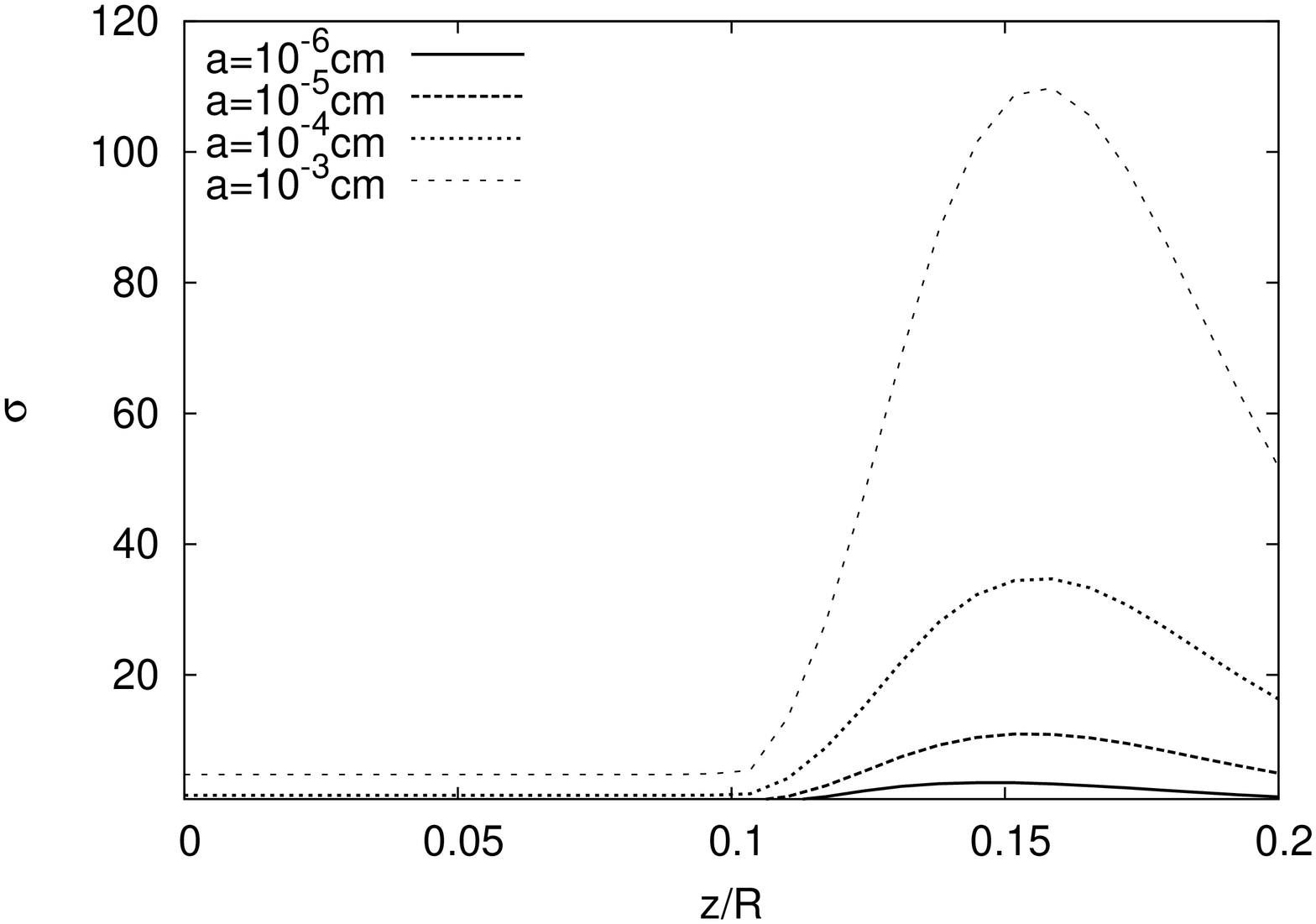} 
\caption{Vertical profiles of the mean charge (upper panel) and the standard deviation of the charge (lower panel) at a distance of 10 AU from the star
(for various grain sizes $a$). }
\label{Fig_charge}
\end{figure}
%% Fig 9 (Q_full_gray and sigma_full_gray) was moved to the end of manuscript
Upper panel of Fig.~\ref{Fig_charge} presents the charge
vertical profile for grains of various sizes at a distance
of 10 AU from the star. The mean grain charge
vanishes at a height $z/R\approx0.1$. Since the dispersion
of the charge is non-zero (see lower panel of Fig.~\ref{Fig_charge}), grains
with charges of both signs are present in this area
(point B). Thus, electrostatic enhancement of the grain growth will operate in the vicinity of lines of zero
charge in the disk.

Determining the charge of grains in the disk atmosphere
is directly related to the radiative transfer modeling.
The deeper the UV radiation penetrates into
the disk, the deeper the layer of near-zero charge lies.
This facilitates the growth of grains due to their higher
number density. We adopted a conservative estimate
for the radiation field, and considered only radiation from the central star, taking into account extinction
by grains with standard interstellar parameters (in
accordance with formula (5.96) of~\cite{2005pcim.book.....T}). A number of
factors (grain growth, scattering, background radiation)
can lead to better penetration of radiation into
the disk. Therefore, the mean grain sizes in the disk
atmosphere presented below are lower limits, and the
dust coagulation rates in the disk atmosphere will become higher in a more detailed consideration of the
radiative transfer.

\begin{figure}
%\setcaptionmargin{5mm}
%\captionsetup{singlelinecheck=false,justification=raggedright}
\includegraphics[width=0.49\textwidth]{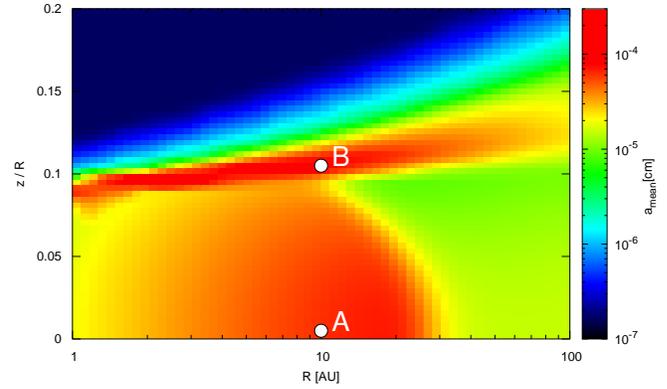} 
\caption{Same as Fig.~\ref{Fig_A_null} for Model 3. The upper boundary of the color scale is two orders of magnitude lower than in Fig.~\ref{Fig_A_null}.}
\label{Fig_A_full}
\end{figure}

\begin{figure}
%\setcaptionmargin{5mm}
%\captionsetup{singlelinecheck=false,justification=raggedright}
\includegraphics[angle=270,width=0.49\textwidth]{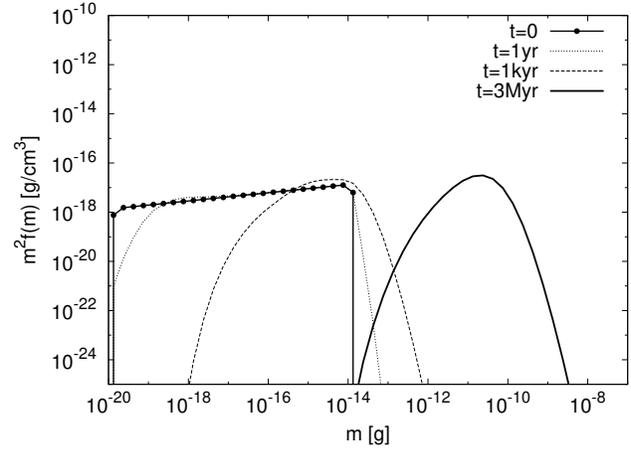} 
\caption{Mass distribution function for the grains at point B for Model 3 (for various times).}
\label{Fig_ff_full}
\end{figure}

\begin{figure}
%\setcaptionmargin{5mm}
%\captionsetup{singlelinecheck=false,justification=raggedright}
\includegraphics[angle=270,width=0.49\textwidth]{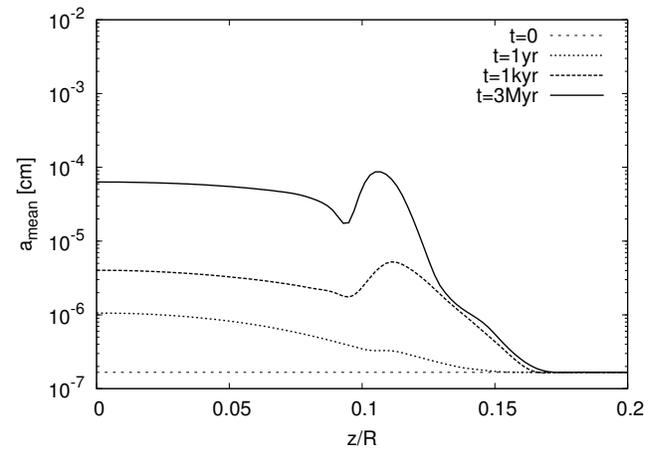} 
\caption{Same as Fig.~\ref{Fig_vertical_null} for Model 3.}
\label{Fig_vertical_full}
\end{figure}

The difference in the densities at points A and B
is approximately three orders of magnitude; i.e., the
coagulation rates for the neutral case differ by six
orders of magnitude. It turns out that this difference
in coagulation rates is fully compensated due to
the absence of Coulomb repulsion and switching to
Coulomb focusing, and leads to an appreciable dust
growth at intermediate heights (Fig.~\ref{Fig_A_full}). The mass
%% Fig 10 (A_full_gray) was moved to the end of manuscript
%Fig.10
%% Fig 10 (A_full_gray) was moved to the end of manuscript
 distribution of the grains at point B (Fig.~\ref{Fig_ff_full}) is not
bimodal, and resembles the distribution for neutral
dust. The dispersion of the charge in early stages ($\sim1$\,kyr) leads to an increase in the dust growth,
even compared to the neutral case. 
%% Fig 11 (ff16_full_gray) was moved to the end of manuscript
%Fig.11
%% Fig 11 (ff16_full_gray) was moved to the end of manuscript
 The grains in the
layer with near-zero charge are even somewhat larger
than those in the dense midplane regions (Fig.~\ref{Fig_vertical_full}).
%% Fig 12 (vertical_full_gray) was moved to the end of manuscript
%Fig.12
%% Fig 12 (vertical_full_gray) was moved to the end of manuscript

This increase in the coagulation rate at the zero-charge
line is a fully expected result. However, in
our formulation, the mean charges of grains of all
sizes vanish at the same zero-charge line, and not at
various regions of space. This makes the appreciable
thickness of the layer of active dust growth that
comes about due to the dispersion of the dust charge an intriguing result. An additional contribution to
the dispersion could also be made by differences in
the work function for the emission of electrons from
grains of inhomogeneous chemical composition. The
absence of an electrostatic barrier at intermediate
heights could play a determining role for the overall growth of dust for a disk that is undergoing sufficiently
strong mixing. A simultaneous description
of mixing processes and the dust growth requires
substantially more complex modeling, and is an interesting
topic for a separate study.

 \section{DISCUSSION}
\label{neUV}  
\subsection{Influence of Dynamical Effects}

The results presented in the previous section were
obtained under the assumptions that the disk structure
is stationary over scales of several million years,
and that there is no drift of the dust relative to the
gas. Turbulence, meridional circulation~\citep{1984SvA....28...50U}, and
sedimentation of the grains will lead to flows into and
out of the layer of active dust growth. The question is
whether these factors facilitate or hinder dust growth.
The answer to this question could be given by appropriate
modeling, but the qualitative picture can also
be described based on general reasoning.

On the one hand, the flow of gas through the
growth layer reduces the residence time for a specific
grain in the layer. This should reduce the coagulation
rate, although the same material may pass through
the growth zone multiple times in the presence of
global vortices or circulation in the disk. On the other
hand, since small grains are efficiently dynamically
coupled to the gas, the flow of gas delivers small
grains to the near-zero charge layer. Due to the lower
charge inside this layer, the coupling between the dust
and gas falls, and the grains are decelerated. They
begin to grow efficiently in this layer, which further
reduces the friction between the gas and the dust per
unit dust mass. Thus, the residence time of the dust
in the layer grows. In other words, a stationary flow
through the growth layer can lead to an increase in
the dust density inside the layer, which enhances the
coagulation rate.

As the grains grow, settling toward the midplane
under the action of the vertical component of the
gravitational force becomes important. The scale
height, determined by the balance between gravitation
and turbulent mixing, is smaller for larger grains.
Other things being equal, the scale height for charged
dust is larger than the scale height for neutral dust,
due to the more effective coupling of charged grains
with the gas. Thus, the charge of a grain increases
the sedimentation scale height, which also enhances
coagulation in the disk atmosphere.

The median lifetime of a protoplanetary disk is
three million years~\citep{2011ARA&A..49...67W}. Disks around massive stars
dissipate more rapidly than disks around solar-type
stars. Therefore, the lifetimes for individual disks vary
from one to ten million years. At late stages of the
evolution, the dispersal of the disk occurs from inside
out, shifting the inner edge of the disk beyond 1 AU
from the star. In Models 1--3, the inner edge of the
disk lies approximately at 0.5 AU, which corresponds
to the dust sublimation radius. The stellar radiation
is efficiently absorbed due to the high densities near
the midplane, and does not reach distances of 1 AU,
so that grains at height $z=0$ are negatively charged.
The dispersal of the disk means that radiation begins
to penetrate into the disk midplane, which changes
the sign of the dust charge. In this way, the layer of
active dust growth (in the region of near-zero charge)
moves outward along the equatorial plane, with its
relatively high number density of dust grains. Since
an increase in grain size leads to a decrease in opacity,
this process will proceed in an accelerating regime.
Therefore, the growth of dust in the region of near-zero
charge can become even more important at the
stage of the dispersal of the disk.

\subsection{Influence of Porosity and Fluctuations
of the Grain Velocities}

The results presented here were obtained for compact,
spherical dust grains. However, to all appearances,
grains in protoplanetary disks have a very
porous, fractal structure. This porosity increases the
effective cross section of the grains, together with
the sticking probability. Therefore, the presented
mean grain sizes are lower limits for the more general
case of non-compact agglomerates. Note that it was
shown in~\cite{2009ApJ...698.1122O} that the charge of grains increases with
their porosity.

The use of the mean relative velocity for grains
of two specified masses can distort the real pattern
of the dust evolution. For example, considering the
velocity distribution of grains of a single size makes
it possible to lower the fragmentation barrier against
dust growth~\citep{2012A&A...544L..16W}. As the grain size increases, the relative
velocity increases, and may exceed the velocity
at which fragmentation occurs. However, grains with
lower velocities will always be found in the ensemble,
whose collisions do not lead to fragmentation. Consideration
of the dust charge provides an additional
possibility for suppressing fragmentation. Since the
grains in the midplane have the same charge sign,
Coulomb repulsion slows particles approaching each
other, lowering the fragmentation rate.

\subsection{Influence of the UV Excess and Degree
of Ionization}

The magnitude of the UV excess determines how
deep the radiation penetrates into the disk, i.e., the
position of the zero-charge line. The more intense
the UV excess, the faster the dust coagulation rates,
since the growth layer moves deeper into the disk. 

We took the degree of ionization to be constant in
the disk, and equal to $x_{\rm e}=10^{-4}$, corresponding to
 ionized carbon. The range of physical conditions
in a protoplanetary disk is very broad, and the degree
of ionization varies widely, from $\sim 10^{-10\div-15}$ to
unity~\citep{2004A&A...417...93S}. However, we used the degree of ionization
only to compute the grain charges. The dust
charge in the disk atmosphere is determined by the
photoelectric effect, and is practically independent of $x_e$. The situation is the opposite in the midplane,
where collisions with electrons and ions determine
the dust charge. However, under the conditions of
electrical neutrality, the dependence on the number
density disappears, and the grain charge depends only
on the ratio of the masses of the dominant ion and the
electron~\citep[formula (5.70)]{2005pcim.book.....T}. Thus, the degree of ionization
is important only in the transition layer, where
carbon is ionized and $x_{\rm e}\sim10^{-4}$. Note that the degree
of ionization in deep layers of the disk could be so
low that grains accrete all available electrons. Their
charges will be smaller in magnitude in this case. The
conclusion that the coagulation rate of the dust grains
is substantially suppressed in the midplane of the disk
is preserved in a self-consistent computation of the
grain charge and degree of ionization~\citep{2011ApJ...731...96O}.

\section{CONCLUSION}

We have considered one problem in the modern
theory of planet formation~--- the electrostatic barrier
against the early growth of dust in protoplanetary
disks. Studies of the coagulation of charged particles
have often assumed that dust grains acquire charge
only via collisions with ions and electrons. Here, we
have conducted modeling of grain growth taking into
account radiative and collisional mechanisms for the
acquisition of charge.

Consideration of the photoelectric effect leads to
the appearance of a line of zero charge at intermediate
heights in the disk (where $A_v\sim1$). The non-zero
dispersion of the grain charge gives rise to a fairly
thick layer in which grains with both signs of charge
are present. This removes the electrostatic barrier in
a certain layer of active dust growth at intermediate
heights, and opens the possibility of removing this
barrier completely for a disk with efficient mixing.
Global flows in the disk, vortices, meridional circulation,
and turbulence should facilitate an inflow of
grains into the growth region from other parts of
the disk. In this case, evolved dust that has passed
through the growth region will return to other areas
with significant electrostatic barriers. The circumstance
that the scale height for the sedimentation of
charged grains is larger than the corresponding scale
height for neutral grains, due to the higher friction
with gas, is important here.

If mixing mechanisms in the disk are not sufficiently
strong, the proposed mechanism for removing
the electrostatic barrier will not operate efficiently in
the stationary stage of evolution of the disk. However,
the zero-charge line will pass through the disk midplane,
where the number density of dust and the dust
coagulation rate are high, in the stage of the dispersal
of the disk. In this regime, the evolution of the dust
can proceed in an accelerating fashion on short time
scales, due to the positive feedback between dust
growth and irradiation in the disk.

\section*{Acknowledgments}

We thank Ya.N. Pavlyuchenkov, N.N. Chugai, and
the anonymous referee for useful comments.

This work was supported by the Russian Foundation
for Basic Research (nos. 13-02-00138, 14-02-
31400), the “Dinasty” Foundation, and St. Petersburg
State University (grant no. 6.38.669.2013).

\bibliographystyle{mn2e} 
\bibliography{paper}

\end{document}